\documentclass[conference]{IEEEtran}
\pdfoutput=1 
\IEEEoverridecommandlockouts
\usepackage[utf8x]{inputenc}
\usepackage{graphicx}
\usepackage{babel,blindtext}
\usepackage{booktabs}
\usepackage{multirow}
\usepackage{siunitx}
\usepackage{latexsym}
\usepackage{array}
\usepackage{url}
\usepackage{float}
\usepackage{subfigure}
\usepackage{times}
\usepackage{commath}
\usepackage{makecell}
\usepackage{caption}

\usepackage{adjustbox}
\usepackage[ruled,vlined]{algorithm2e}
\usepackage[table,xcdraw]{xcolor}
\usepackage[section]{placeins}
\usepackage{amsmath,amsfonts,amsthm,bm} 
\DeclareMathOperator*{\argmax}{argmax} 
\newcolumntype{P}[1]{>{\centering\arraybackslash}p{#1}}

\title{GANash - A GAN approach to steganography}

\begin{document}

\author{\IEEEauthorblockN{ Venkatesh Subramaniyan}
\IEEEauthorblockA{\textit{SSN College of engineering} \\
Chennai, India \\
venbiseven77@gmail.com}
\\
\IEEEauthorblockN{Vagheesan A K}
\IEEEauthorblockA{\textit{SSN College of engineering}\\
Chennai, India \\
vagheesanak@gmail.com }
\and
\IEEEauthorblockN{ Vignesh Sivakumar}
\IEEEauthorblockA{\textit{SSN College of engineering} \\
Chennai, India \\
vignesh25sivakumar@gmail.com } \\
\IEEEauthorblockN{Dr. K.J Jegadish Kumar}
\IEEEauthorblockA{\textit{Associate Professor, ECE Department} \\
\textit{SSN College of engineering} \\
Chennai, India \\
jegadishkj@ssn.edu.in }
\and
\IEEEauthorblockN{S.Sakthivelan}
\IEEEauthorblockA{\textit{SSN College of engineering} \\
Chennai, India \\
sakthissv1999@gmail.com }\\
 \IEEEauthorblockN{Dr. K.K.Nagarajan}
\IEEEauthorblockA{\textit{Associate Professor, ECE Department} \\
\textit{SSN College of engineering} \\
Chennai, India \\
nagarajankk@ssn.edu.in  }

}

\date{}
\maketitle
\begin{abstract}
Data security is of the utmost concern of a communication system. Since the early days, many developments have been made to improve the performance of the system. PSNR of the received signal, secure transmission channel, quality of encoding used, etc. are some of the key attributes of a good system. To ensure security, the most commonly used technique is cryptography in which the message is altered with respect to a key and using the same, the encoded message is decoded at the receiver side. A complementary technique that is popularly used to insure security is steganography. The advancements in Artificial Intelligence(AI) have paved way for performing steganography in an intelligent, tamper-proof manner. The recent discovery by researchers in the field of Deep Learning(DL), an unsupervised learning network known as the Generative Adversarial Networks(GAN) has improved the performance of this technique exponentially. It has been demonstrated that deep neural networks are highly sensitive to tiny perturbations of input data, giving rise to adversarial examples. Though this property is usually considered a weakness of learned models, it could be beneficial if used appropriately. The work that has been accomplished by MIT for this purpose, a deep-neural model by the name of SteganoGAN, has shown obligation for using this technique for steganography. In this work, we have proposed a novel approach to improve the performance of the existing system using latent space compression on the encoded data. This theoretically would improve the performance exponentially. Thus, the algorithms used to improve the system's performance and the results obtained have been enunciated in this work. The results indicate the level of dominance this system could achieve to be able to diminish the difficulties in solving real-time problems in terms of security, deployment and database management.
\end{abstract}


\section{\bfseries Introduction}
\label{Intro}
Since World war II, encryption has played a very crucial role in secure transmission of data. The two most popular techniques used to transmit data securely over a channel are cryptography and steganography. These are comparable to the two sides of a coin where the former inconceivably encrypts the message while the latter intends to hide the traces of a communication itself. In older days, these techniques were done manually but since the arrival of the digital era, computers have been used to do the same. Cryptography has improved drastically over the years, but steganography has been used comparatively less and there is scope for significant improvements.\par
Steganography could be classified primitively on the basis of the type of message used which includes text, images, audio and video. It could be further classified based on the technique used namely spatial domain techniques and transform domain techniques. These techniques employ statistical approaches to hide a message which could be easily detected by intruders. The digital age has led to the discovery of Neural Networks which tries to mimic human knowledge to be able to solve real-time problems in unanticipated ways. The same has been used to perform steganography to overcome the disadvantages of the conventional methodologies. Researchers have discovered and developed algorithms in this regard. Thus, the intrications of neural networks have been understood from their primitives.\par
Advancements in neural networks have led to the discovery of Generative Adversarial Networks(GANs) \cite{goodfellow2014generative}. GAN is a category of machine learning framework in which two neural networks compete against each other to achieve the specified objective. GAN is the most commonly used steganography technique in recent times. GAN has been efficient in achieving excellent results in various fields. One such field is steganography where the induction of GAN has proved to be authoritative compared to the conventional algorithms. The most recent work using GAN for performing steganography has been accomplished by researchers at MIT known by the name of SteganoGAN \cite{zhang2019steganogan}. This model shows promise in terms of security with the ability to hide more amount of data in a given cover image compared to the conventional techniques. Another popular work known as Deep steganography \cite{NIPS2017_838e8afb} also adds value to the same. Thus, they are effective in terms of security.\par
The major downside of the above mentioned systems is the computation cost. The applications of steganography not only comprises secrecy but also expands to various other things such as protection of data alteration, data storage, digital content distribution, etc. In such cases, the technique needs to be computationally effective to be used in real-time. A well designed steganographic network must be able to withstand great amount of intrusions. Thus, the system has to be secure and also deployable. In this work, a lightweight model is built using GAN to perform image steganography with latent space compression. The main objective of this work is to improve the performance in terms of time taken to encode and decode, which is imperative in case of deployment. The model has been named GANash since GAN is used to perform steganography on a given data and achieve a state of Nash equilibrium.\par

\section{\bfseries Related Work}\label{related}
There are various methods of implementing the technique of Image steganography. The 3 major types of image steganography techniques used are - Conventional Steganography, Convolutional Neural Network(CNN) based steganography and Generative Adversarial Network(GAN) based steganography. In this section the various algorithms and architectures used are discussed.
\subsection{Conventional Steganography} Conventional steganography is an old technique which involves bit manipulation of images, that is, the pixel values are altered to embed the information. These manipulations can either be a wavelet transform, a simple Least Significant Bit (LSB) or a bitwise XOR operation.
They involve basic spatial domain or frequency domain operation in order to output steganographic images.\par
\textbf{Neeta D.et al(2006)} \cite{lsb} proposed a steganography method where the Least Significant Bit(LSB) gets manipulated by bitwise OR operation. This method can be used efficiently for images with 8 bits and 24 bits. According to this method, by changing two least significant bits, the human eye perception could be deceived.
\textbf{Yuan-Hui Yu et.al} \cite{lsbGrey} proposed a method where a true color image can be embedded with grey scale and color image. The embedded image data are then subjected to standard DES cryptography algorithm. This methodology proved to be better compared to \cite{lsb}.
\textbf{Al Atabi et.al} \cite{wavelet} proposed a method which uses wavelet transform. In this method, the information is translated with respect to wavelet transform coefficients. \par These methods are conventional in nature and are mathemathical based transforms, which might be efficient to some extent but are easily detectable by statistical analysis and could get unveiled easily. Thus, the robustness of neural networks comes into play in which the bit manipulation is quite hard to track.
\subsection{Convolutional Neural Network based Steganography}
In this type of steganography, the ability of neural network, a sophisticated mathematical model, is used. One such type of network that is popularly used on images is the Convolutional Neural Network(CNN), which uses convolution operation with the help of kernels(filters) to extract and compress the information present in the original image.\par

\textbf{Pin Wu et.al} \cite{stegnet} proposed a steganography method which utilizes both deep convolutional neural network and image-image translation \cite{cyclegan}. In this approach the end-to-end mapping between cover and embedded image, hidden image and decoded image could be learnt. Hence this method is more robust and has high capacity compared to the traditional ones.
\textbf{Tang et.al} \cite{advEmb} presented a novel method of steganography which used an approach called adversarial embedding(ADV-EMB) that could accomplish the criteria of hiding a stego message while at the same time surviving a convolutional neural network (CNN) based steganalyzer.\par
The above works discuss only about steganography or the process of embedding data but the work accomplished by \textbf{Qian Y et.al} \cite{dlSteg} discusses about the process of steganalysis using deep learning models.  The proposed model  has the capability to learn complicated features in the image, which  steganalysis uses to detect steganography. Contrasted with the current schemes ,Just with a few convolutional layers this model can easily learn complex dependencies. Since the process of feature extraction and classification are carried under the single architecture,the knowledge of classification can be utilized in feature extraction as well. Hence this learnt feature can be used to identify whether the steganography has been performed or not.

\subsection{GAN based Steganography}
Generative adversarial network is a class of neural networks where two neural networks compete to minimise the loss. Due to its generative and discrimininative properties, it finds a wide range of applications which also includes steganography. \par
\textbf{Volkhonskiy et al.} \cite{volkhonskiy2019steganographic} proposed a new model for generating image-like cover based on Deep Convolutional Generative Adversarial Networks (DCGAN)\cite{dcgan} and this method is secure from steganalysis. In this network architecture, a generative model is trained for image stego-cover by challenging it with two deep convolutional adversaries: a discriminator network, which optimizes the output to look like samples from the real dataset, and a steganographic analyzer, which aims at detecting if an image conceals a hidden message.\par
\textbf{Shi et.al} \cite{shi2018ssgan} proposed a novel strategy of Secure Steganography based on Generative Adversarial Network. The proposed architecture has one generative network, and two discriminative networks. The generative network essentially assesses the visual nature of the generated images from steganography, and the discriminative network are utilized to assess their appropriateness for data hiding.\par
 \textbf{Zhang et.al} \cite{zhang2019steganogan} proposed a novel procedure for concealing binary information in images using generative adversarial network. This technique improves the cognitive nature of the images created by this model and accomplishes best in class payloads of 4.4 bits per pixel. In addition, this also dodges discovery by steganalysis tools and is compelling on images from various datasets.\par

\section{\bfseries Generative Adversarial Networks}
Generative Adversarial Network(GAN) is a class of machine learning framework invented by Ian Goodfellow and his colleagues in 2014 \cite{goodfellow2014generative}. In this architecture, two neural networks try to reach a common objective by treating each other as an adversary. Given a training set, this technique learns to generate new data with the same statistics as the training set. GAN uses semi-supervised learning to learn any kind of data distribution. It aims at learning the true data distribution $p_{real}$ of the training records so as to generate synthetic data records with some minimal variations in the learnt distribution $p_g$ while the discriminator tries to assign the probability score for the target $P(Y|X)$  with the given training samples. GANs are formulated as a minimax game, where the Discriminator is trying to minimize its reward $V(D, G)$ and the Generator is trying to maximize its loss. It can be mathematically described using the formula 
\begin{equation}
    V(D,G) = E_{x\sim p_{data}(x)}[logD(x)] +E_{z\sim p_{z}(x)}[log(1-D(G(z))] 
\end{equation}
\subsection{Nash-equilibrium}
The intention of GAN is to minimize the divergence caused by true and learnt data distributions. Consider two distribution functions $f$ and $g$. Assume $g$ is the generated distribution from $f$ and $D$ is a divergence function. $D$ takes $f$ and $g$ as inputs and measures the divergence value and $D$ satisfies the condition $D(f,f)=0$.
\begin{equation}
    \min_{\theta}(f_{0},g_{\theta})\label{eq}
\end{equation}
Equation \eqref{eq} shows the optimization objective, where $f_0$ is real distribution with fixed parameters and $g_\theta$ is the generated distribution with $\theta$ parameters. Adapting SimGA from \cite{mescheder2018numerics}, the local Nash-equilibrium point $(\bar\psi,\bar\theta)$ is derived by
\begin{equation}
\begin{aligned}
    \bar\psi \in \argmax_\psi f(\psi,\bar\theta) \\
    \bar\theta \in \argmax_\theta f(\theta,\bar\psi)
\end{aligned}
\end{equation}

\section{\bfseries Data pipeline}
 In this section, the data pipeline and the design choices for optimizing the performance is discussed.
\subsection{Pre-processing data pipeline}
The main focus here is on reducing the RAM over-usage while designing the pre-processing pipeline. It is known that CPU cores can perform on different instructions (MIMD) in parallel at a given time. This is leveraged in the below stated pipeline.
Some of the standard computer architecture aspects to be considered beforehand for getting the best results are :
\begin{enumerate}
    \item CPUs are optimized based on memory access time(latency optimization) while GPUs are optimized based on bandwidth.
    \item CPUs can carry out optimized and sophisticated tasks whereas GPUs could face bottleneck issue i.e. transferring a huge amount data to the GPU is time consuming.
\end{enumerate}
Based on these standards, the data pre-processing is performed on the CPU and model's mathematical computations are performed on the GPU.
\subsection{Datasets}
\subsubsection{DIV2K}
It is a novel 2K resolution benchmarking data set \cite{8014884}. This data set is chosen for its high resolution details and its large diversity of contents including people, handmade objects and environments such as flora and fauna.
\subsubsection{MS COCO}
It is a state-of-the-art object recognition data set \cite{lin2015microsoft} which consists of everyday scenes containing common objects and these multi class object data sets can improve GANash's stego abilities among realistic image
environment.
\subsection{Text to bits conversion}
Text to bits converstion is adapted from \cite{zhang2019steganogan}. Reed-Solomon technique \cite{Reed1960PolynomialCO} has been proved to be effective in error correction while translating from decoded message bits into text. One disadvantage in using this technique is that it introduces data redundancy in message bits.
\subsection{Parallelization}
Tensorflow data API is used to perform parallel level transformation. This API enables the pipeline to choose the number of threads needed for data level mapping. The addresses of the training images is loaded and only a mini-batch of it will be selected as load into the pipeline. Because of this, only a small amount of RAM is occupied during each iteration. Hence, this efficiently improves the performance of the pipeline.

\section{\bfseries Model Architecture}
In this work, a GAN based model with an encoder, a decoder and a critic has been proposed. In this section, the design choices for GANash and its loss functions are discussed in detail.
\subsection{Critic}
\begin{figure}[htbp]
\centerline{\includegraphics[width=90mm,height=50mm]{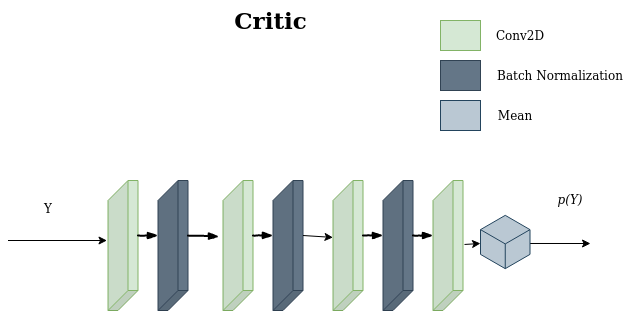}}
\caption{Critic Network}
\label{fig:Critic}
\end{figure}

The architecture of the critic network is shown in fig \ref{fig:Critic}. It consists of 4 stages. The first 3 stages include a series of convolutional layer with $3 \times 3$ kernels, a leaky relu layer and followed by a batch normalization layer. The last stage is a convolutional layer with $1 \times 1$ kernels and is further reduced to its average mean $p(Y)$.\par
 The Critic network tries to maximize the convergence between the two distributions. Let $R_0$ represent the real distribution and $E_\theta$ represent the encoder distribution. The critic network helps to minimize the divergence of $E_\theta$ with $R_0$ using \eqref{eq}. It takes $Y$ as input, which could either be the cover image $I$ or the stego image $S$ and returns the probability score $p(Y)$.

\subsection{Encoder}
\begin{figure}[H]
\centerline{\includegraphics[width=90mm,height=50mm]{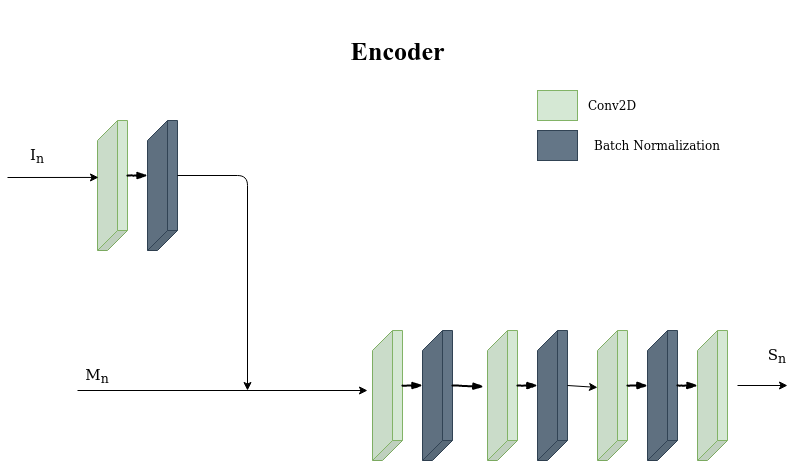}}
\caption{Encoder Network}
\label{fig:Encoder}
\end{figure}

 The encoder architecture consists of 4 stages as shown in fig \ref{fig:Encoder}. The first 3 stages consist of a convolutional layer with $3 \times 3$ kernels followed by a batch normalization layer. The final stage is a convolutional layer consisting of $3 \times 3$ kernels with tanh activation.\par

Based on the critic network's feedback, the encoder learns to hide the message within the cover image and accordingly it changes the parameters $\theta$ of its distribution $E_\theta$. It takes Input image $I_n$, Message to hide $M_n$ and using the trained parameters $\bar\theta$, it generates the stego image $S_n$.

\subsection{Decoder}
\begin{figure}[htbp]
\centerline{\includegraphics[width=90mm,height=50mm]{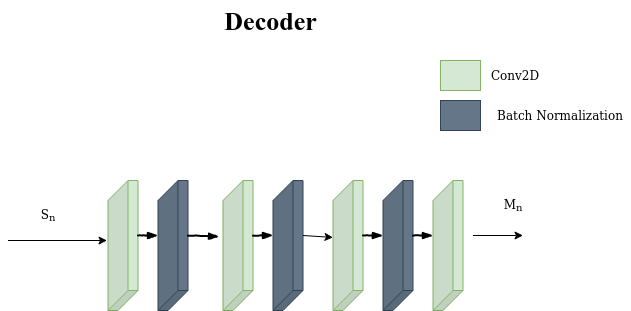}}
\caption{Decoder Network}
\label{fig:Decoder}
\end{figure}

The decoder architecture also consists of 4 stages as shown in fig \ref{fig:Decoder}. The first 3 stages consist of a convolutional layer with $3\times3$ kernels followed by a batch normalization layer. The final layer is a convolutional layer with kernel size $3\times3$ and number of filters equal to data depth.

The decoder network learns to unravel the message from the stego image. Based on the data depth D, the final layer of the decoder should be changed to B$\times$H$\times$W$\times$D, where B is the mini-batch size, H and W are the height and width of the cover image respectively. It takes $S$ as input and returns the message $M$. 

In addition to this, a global steganographic model can be prepared by pre-training GANash on higher data dimension D, so that the model can be fine-tuned for varying dimensions.
\subsection{Loss functions} \label{loss_section}
The loss functions for encoder, decoder and critic are given in \eqref{loss}. Mean Squared Error(MSE) is used to calculate the encoder and the critic loss. Sigmoid Cross-Entropy(SCE) is used to calculate the decoder loss. 
\begin{equation}
\label{loss}
\begin{aligned}
&    l_{enc}= \frac{1}{N}\sum_{i=1}^{N} (I-S)^2 \\\\
&    l_{dec} = -\sum D(M).log(D(M)) \\\\
&    l_{critic}= MSE[p(S)-p(I)] \\\\
& l_T  = l_{enc} + l_{dec} + l_{critic}
\end{aligned}
\end{equation}

where $I$, $S$ and $M$ are the input image, stego image and message respectively and $D$ represents the decoder objective function.
Section \ref{marker} briefs the process of adversarial training with these loss functions.
\subsection{Training}\label{marker}
The model is trained on a 12 core CPU machine with GTX 1050ti-16GB GPU. The encoder-decoder network and the critic network are iteratively optimized using the loss functions presented in section \ref{loss_section}. The objective of this training is to maximize the convergence between true and fake distributions. This is ensured by the following three steps :
\begin{enumerate}
    \item Critic is optimized based on the probability scores on the two distributions until it stations the model at Nash-equilibrium .
    \item Decoder is optimized depending on the decoder loss $l_d$.
    \item Finally, the gradients of the encoder, decoder and critic variables are updated based on the total loss $l_T$.
    
\end{enumerate}

\begin{algorithm}[htbp]
\SetAlgoLined
\KwResult{ $E_\theta : I \mapsto S$, Given $ \min_{\theta}(f_{0},g_{\theta}) \mapsto 0$}
 Initialize\;
 \textbf{Require}: total loss $l_t$ and decoder loss $l_d$\;
 \While{not converged}{
  1. Compute gradient for Critic\;
  $x_\psi\leftarrow\nabla_\psi R(\theta; \psi)$ \;
  $x_\theta\leftarrow\nabla_\theta E(\theta; \psi)$ \;
  Update the variables\;
  $\theta\leftarrow\theta + \eta x_\theta$ ; $\psi\leftarrow\psi + \eta x_\psi$ \;
  2. Compute gradient for Decoder\;
  $x_d\leftarrow \nabla_d D(l_d;d)$ \;
  Update the variables\;
  $d \leftarrow d + \eta x_d$ \;
  3. Overall model optimization \;
  $x_{e,d,c} \leftarrow \nabla_{e,d,c} T(l_t;e,d,c)$ \;
  update encoder $e$, decoder $d$ and critic $c$ variables 
 }
 \caption{Triplet training}
\end{algorithm}

\subsection{Hyperparameters}
The parameters are set on the basis of experimentation. The final set of hyper-parameters used in training the GAN with which the required results are obtained is given in table \ref{tab:params}.

\begin{table}[H]
\centering
\caption{Hyperparameters}
\label{tab:params}

\begin{tabular}{|l|P{1.2cm}|}
    \hline
    \thead{\bfseries parameters} & \thead{\bfseries value}  \\
    \hline
    Learning rate - Critic  & 1e-5  \\
    Learning rate - Decoder &  1e-2\\
    Learning rate - Total & 1e-5 \\
    Gradient clip rate & (-0.1,0.1) \\
    Hidden dims & 32 \\
    Data depth & [3,4,5] \\
    Kernel size & 3 \\
    Coworkers & 4 \\
    Buffer & 8 \\
\hline
\end{tabular}
 
\label{tab:dataset}
\end{table}
\section{\bfseries Evaluation metrics} \label{eval_metrics}
The model is evaluated based on the standard metrics used for a steganography system. The metrics used are listed below.\par
\subsection{Payload}
It is defined as the ratio of the number of message bits to the size of the cover image.

\begin{equation}
    Payload=\frac{Number\;of\;bits\;in\;message}{Size\;of\;the\;cover\;image}
\end{equation}
where size of cover image = H$\times$W.

\subsection{Time To Enocde (T2E)}
It is the time elapsed to encode the message into cover image to obtain the stego image.

\subsection{Time To Decode (T2D)}
It is the time elapsed to decode the stego image to obtain the message.

\subsection{Mean Square Error (MSE)}
It is defined as the average squared difference between the cover and stego image.

\begin{equation}
    MSE=\sum_{j=1}^{H} \sum_{k=1}^{W} \frac{[C(j,k)-S(j,k)]^2}{H\times W}
\end{equation}
where C is Cover image, S is Stego image.

\subsection{Peak Signal To Noise ratio (PSNR)}
The percentage of noise present in the stego image is defined as the Peak Signal to noise ratio.

\begin{equation}
    PSNR=10log_{10}  \frac{(2^n-1)^2}{MSE}
\end{equation}
where n is the number of bits used to represent a pixel in the cover image.

\subsection{Cross-Correlation Coefficient (r)}
It is a metric that compares the similarity between the cover image and the stego image.

\begin{equation}
    r = \frac{(C-m_1)(S-m_2)}{\sqrt{(C-m_1)^2(S-m_2)^2}}
\end{equation}

where $m_1$ and $m_2$ are the mean of the cover and stego image respectively.

\subsection{Security}
It is a measure of immunity of the encoding technique against intrusion.
\section{\bfseries Results and Discussion}
\begin{figure}[H]
    \centering
    \subfigure[]
    {
        \includegraphics[width=1.4in]{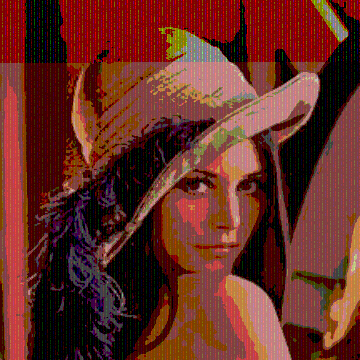}
        \label{fig:first_sub}
    }
    \subfigure[]
    {
        \includegraphics[width=1.4in, height = 3.565cm]{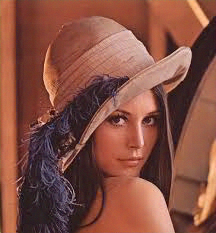}
        \label{fig:second_sub}
    }
    \subfigure[]
    {
        \includegraphics[width=1.4in]{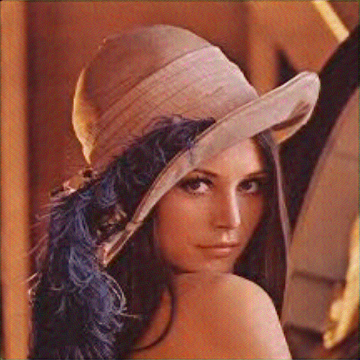}
        \label{fig:third_sub}

    }
    \captionsetup{justification=justified,margin=0.1cm}
    \caption{Qualitative validation of the stego image using (a)LSB with a payload of 2.315 bits/pixel (b)SteganoGAN with a payload of 4 bits/pixel (c)GANash with a payload of 4 bits/pixel}
    \label{fig:Lena}
\end{figure}

This section showcases the results obtained using the various evaluation metrics listed in section \ref{eval_metrics}. The cover image used for hiding the message to obtain the stego image is that of Lenna Forsén. The dimension of the image is $360\times360$ thus containing a total of 1,29,600 pixels. Fig \ref{fig:Lena} shows the stego image obtained using the various techniques listed in this paper which are LSB, MIT's Steganogan and the model implemented through this work, GANash.

\begin{table}[H]
\centering
\caption{Quantitative result}
\begin{tabular}{|p{3.3cm}|P{1cm}|P{1.6cm}|P{1.2cm}|}
 \hline
 \thead{\bfseries Metric} & \thead{\bfseries LSB} & \thead{\bfseries SteganoGAN} & \thead{\bfseries GANash}\\
 \hline
Max. payload (bits/pixels) & 1 & 4 & 4\\
 
Time to Encode (secs) &   3.56  & 1.968   & 0.129\\
 
Time to Decode (secs) & 2.3  & 46.12 &  0.174\\
 
Mean Squared Error(MSE)  & 0.00237 & 0.00304 &  0.000361\\

PSNR (dB) & 77.375  & 76.28 & 85.5410\\

Cross-Correlation coefficient & 0.98684  & 0.98731   &0.99326\\

Security& low  & high & high\\
 \hline

\end{tabular}
\end{table}
From the results obtained, it is evident that GANash and SteganoGAN are superior compared to LSB. In addition, GANash has proved to be effective compared to SteganoGAN in terms of Time to encode and decode, Mean Squared Error and  PSNR.

\section{\bfseries Conclusion}
In this work, a lightweight, cost-effective, intelligent model has been introduced to perform steganography in a robust and automated manner that is compatible with low specification computing engines. Further, our model is trained in a way that the encoding and decoding parts are distinguishable without any loss in the training information. In addition to this, we efficiently handled our model training to concentrate more on lossless message decoding. It is evident from the results obtained that the model presented in this work is ten times more powerful compared to the existing model.

\section*{Acknowledgements}
We would like to thank Raviteja Vuppaladhadiyam and Harine Govindarajan from Sri Sivasubramaniya Nadar College of Engineering for assisting and proofreading the work.

\bibliographystyle{IEEEtran}
\bibliography{ncictgan}

\end{document}